\documentclass[aps,prb]{revtex4}
\usepackage{graphicx}
\usepackage{subfigure}  
\usepackage{amsmath} 
\usepackage{amssymb}
\begin{document}
\newcommand{\pderiv}[2]{\frac{\partial #1}{\partial #2}}
\title{Interplay between field-induced and frustration-induced 
quantum criticalities in the frustrated two-leg Heisenberg ladder}
\author{Brandon W. Ramakko}
\author{Mohamed Azzouz} 
\email[Electronic Address: ]{mazzouz@laurentian.ca} 
\affiliation{Laurentian University, Department of Physics,
Ramsey Lake Road, Sudbury, Ontario P3E 2C6, Canada.}

\date{August 21, 2007}

\begin{abstract}
The antiferromagnetic Heisenberg two-leg ladder
in the presence of frustration and an external magnetic field is a system that 
is characterized by two sorts of quantum criticalities, not only one.
One criticality is the consequence of intrinsic frustration, and the 
other one is a result of the external magnetic field.
So the behaviour of each of them in the presence of the other
deserves to be studied. Using the 
Jordan-Wigner transformation in dimensions higher than one 
and bond-mean-field 
theory we examine the interplay between the field-induced and 
frustration-induced quantum criticalities in this system. 
The present work could constitute 
a prototype for those systems showing 
multiple, perhaps sometimes competing, quantum criticalities.
We calculate several physical quantities like the 
magnetization and spin susceptibility 
as functions of field and temperature.
\end{abstract}

\maketitle
\newpage

\section{Introduction}

In low-dimensional antiferromagnetic (AF) Heisenberg spin ladders, 
the mixing of magnetic frustration, external magnetic fields, and 
thermal fluctuations together with quantum fluctuations is a recipe 
for interesting non-conventional and exotic behaviours. The two-leg frustrated 
ladder is an example, which is characterized by three distinct 
non-magnetic quantum spin liquid states; the N\'eel-type (N-type) state, 
ferromagnetic-type rung (R-type) state, and ferromagnetic-type chain 
(F-type) state. 
These states are characterized by ferromagnetic spin 
arrangements along the diagonals, rungs, or chains respectively. 
In a recent work, we studied the effect of 
thermal fluctuations on these states and found that the quantum 
phase transitions between the R-type and N-type states on 
one hand and between
the R-type and F-type states on the other hand
evolve with temperature into classical phase transitions 
between these disordered
states~\cite{ramakko2007}. 
Numerical studies~\cite{Jd4,Honecker2000} have shown that in the 
presence of a magnetic field, frustration is able to stabilize 
a magnetization plateau at half the saturation magnetization. 
In terms of the Jordan-Wigner (JW) fermions~\cite{Azz1}, this 
would mean that a plateau in the magnetization appears because 
of the appearance of an energy gap between two (excitation) energy 
bands. In this paper we will  check this claim within the 
analytical bond-mean-field theory 
(BMFT)~\cite{ramakko2007,Azz1,Azz2,Azz3,Azz5}. We will discuss the 
interplay between the quantum criticality induced by frustration
and that produced by magnetic field. Indeed, magnetic field 
has been shown recently to induce quantum criticality
between two distinct states, one characterized by spin bond ``order''
and the other one by no spin bond order, in the Heisenberg chain and 
non-frustrated two-leg ladder~\cite{Azz5}. Note that spin bond order
does not imply any kind of magnetic long range order; i.e., 
all the states are magnetically disordered in agreement with the 
Mermi-Wagner theorem~\cite{mermin1966}.

In section \ref{sec:method} the BMFT is 
briefly reviewed. In section \ref{sec:results} 
the magnetization, phase diagram, mean-field parameters, 
energy spectra, and susceptibility are examined. 
In section~\ref{sec:Conclusion} conclusions are drawn.

\section{Method}
\label{sec:method}

\begin{figure}
		\includegraphics[height=2.20cm]{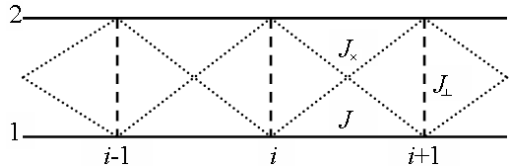}
		\centering
	\caption{The two-leg ladder showing the couplings along the chains, 
		         rungs, and diagonals is displayed.}
	\label{fig:Two leg ladder}
\end{figure}

 We apply BMFT following the method described in Ref.~\cite{Azz5}. 
The Hamiltonian for the spin-$\frac{1}{2}$ two-leg ladder with 
diagonal interactions in a magnetic field is written as
\begin{equation}
\label{eq:Ham}
H = J \sum_{i}^N \sum_{j=1}^2 \textbf{S}_{i,j}\cdot \textbf{S}_{i+1,j} +
J_\perp \sum_{i}^N \textbf{S}_{i,1}\cdot\textbf{S}_{i,2}
+ J_{\times} \sum_{i}^N ( \textbf{S}_{i,1}\cdot\textbf{S}_{i+1,2} 
+ \textbf{S}_{i+1,1}\cdot\textbf{S}_{i,2}) -h \sum_{i}^N 
\sum_{j=1}^{2} S_{i,j}^{z},
\end{equation}
where $J$ is the coupling along the chains, $J_\perp$ the transverse coupling, 
and $J_{\times}$ the coupling along the diagonals as seen in 
Fig.~\ref{fig:Two leg ladder}. The index $i$ labels the position 
of the spins along the two chains, each of which has $N$ sites. 
The first term represents the interactions of nearest-neighboring 
spins along the chains (legs) of the ladder, the second term represents 
the interactions of  the spins along the rungs, and the third term sums 
the interactions along the diagonals. As usual, $\textbf{S}_{i,j}$ is 
the spin operator. Here $h=g\mu_B B$ with $B$ being the magnetic field, 
$g$ the Land\'e factor, and $\mu_B$ the Bohr magneton.
%
%
%

The JW transformation \cite{Azz1,JW} for the two-leg Heisenberg 
ladder is defined as \cite{Azz3}
\begin{eqnarray}
\label{eq:2D JW}
S_{i,j}^{-} & = & c_{i,j}e^{i\phi_{i,j}}, \qquad S_{i,j}^{z}  
=  n_{i,j}-1/2,\qquad 
n_{i,j}=  c_{i,j}^{\dag}c_{i,j},\nonumber \\
\phi_{i,1}\!& \!=\! &\! \pi[\sum_{d=0}^{i-1}\sum_{f=1}^{2}n_{d,f}], 
\qquad \phi_{i,2}  
=  \pi[\sum_{d=0}^{i-1}\sum_{f=1}^{2}n_{d,f} + n_{i,1}].
\end{eqnarray}
Here $i$ and $j$ are the coordinates along the chain and rung 
directions, respectively. 
The phases $\phi_{i,j}$ are chosen so that all the spin commutation relations 
are preserved. The $c_{i,j}^{\dag}$ operator creates a spinless fermion at 
site $(i,j)$, while $c_{i,j}$ annihilates one, and $n_{i,j}$ is 
the occupation number operator. Using~(\ref{eq:2D JW}), 
the Hamiltonian~(\ref{eq:Ham}) becomes
\begin{eqnarray}
\label{eq:Ham1}
H && =  \frac{J}{2} \sum_{i}^N (c_{i,1}^{\dag}e^{i\pi n_{i,2}}c_{i+1,1} 
+ c_{i,2}^{\dag}e^{i\pi n_{i+1,1}}c_{i+1,2} + \textrm{H.c.})  + 
\frac{J_\perp}{2} \sum_{i}^N 
(c_{i,1}^{\dag}c_{i,2} + \textrm{H.c.}) \nonumber \\
&&+ \frac{J_{\times}}{2} \sum_{i}^N (c_{i,1}^{\dag}
e^{i\pi (n_{i,2}+n_{i+1,1})}c_{i+1,2} 
+ c_{i+1,1}^{\dag}c_{i,2} + \textrm{H.c.}) \nonumber
 + J \sum_{i}^N \sum_{j=1}^2 (n_{i,j} 
- \frac{1}{2})(n_{i+1,j} - \frac{1}{2}) \nonumber \\
&& + J_\perp \sum_{i}^N (n_{i,1} - \frac{1}{2})(n_{i,2} 
- \frac{1}{2}) + J_{\times}\sum_{i}^{N} [(n_{i,1} 
- \frac{1}{2})(n_{i+1,2} - \frac{1}{2}) \nonumber \\
&& + (n_{i+1,1} - \frac{1}{2})(n_{i,2} 
- \frac{1}{2})]- h \sum_{i}^N \sum_{j=1}^{2} (n_{i}-\frac{1}{2}).
\end{eqnarray}
After applying the JW transformation the Ising terms are decoupled using 
the Hartree-Fock approximation, which neglects fluctuations 
around the mean field points; 
$(O-\langle O\rangle)(O'-\langle O'\rangle) \approx 0$, 
where $O$ and $O'$ are quadratic 
in $c^\dag$ and $c$ \cite{Azz2}. To apply BMFT we introduce 
three mean-field bond parameters; $Q$ in the longitudinal direction, 
$P$ in the transverse direction, and $P'$ along the diagonal. These 
can be interpreted as effective hopping energies for the JW 
fermions in the longitudinal, transverse and 
diagonal directions, respectively \cite{Azz1}:
\begin{equation}
\label{eq:bondparameters}
Q  =  \langle c_{i,j}c_{i+1,j}^{\dag}\rangle,\qquad P  
=  \langle c_{i,j}c_{i,j+1}^{\dag}\rangle,\qquad P'  
=  \langle c_{i+1,j}c_{i,j+1}^{\dag}\rangle.
\end{equation} 
%
%
We choose to place an alternating phase of $\pi$ along the chains 
so that the phase per plaquette is $\pi$ \cite{affleck1988}. 
This configuration is used 
to get rid of the phase terms in the Hamiltonian. We also set $Q_{i,j} 
= Qe^{i\Phi_{i,j}}$ where $Q$ is site independent \cite{Azz5}. Here 
$\Phi_{i,j}$ is the phase of the bond along the chain such that 
$\phi = \pi$ or $0$. This is necessary in order to recover the proper 
result in the limit $J_{\times}$ and $J_{\perp}$ becoming zero, 
in which we get a result comparable to that of des Cloiseaux 
and Pearson~\cite{Cloiseaux} for the spin excitation spectrum 
for a single Heisenberg chain, $E(k) = \frac{\pi}{2}J\left|\sin k\right|.$

The magnetization per site is 
$M_{z} = \langle S_{i,j}^{z}\rangle=\langle n_{i,j}\rangle-\frac{1}{2}$. 
Once a magnetic field is applied the magnetization must be included 
in the decoupling of the Ising terms. The latter must be decoupled 
as many ways as is physically acceptible in order to account for 
all effects. So in addition to the previous two 
ways~\cite{ramakko2007} it is now decoupled in the following third way:
\begin{equation}
JS_{i,j}^{z}S_{i+1,j}^{z}
\approx  J M_{z} c_{i+1,j}^{\dag}c_{i+1,j} 
+ J  c_{i,j}^{\dag}c_{i,j}M_{z} - J M_{z}^2 - J M_{z}.
\end{equation}
The mean-field Hamiltonian becomes
\begin{equation}
H = \sum_{k}\Psi_{k}^{\dag}\mathcal{H}\Psi_{k}+ 2NQ^2 
+ NJ_{\perp}P^{2} +Nh-N(2J + 2J_{\times} + J_{\perp})M_{z}(M_{z}+1).
\end{equation} 
\begin{figure}
\includegraphics[height=4.8cm]{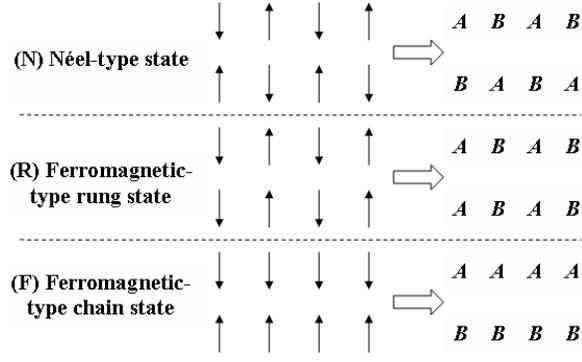}
\centering
\caption{In the left panel, the three possible 
ground states of the system in the Ising limit
are drawn. In the right
panel, the labeling of sublattices corresponding 
to the short-range spin orders that replace the long-range ones 
in the Heisenberg limit are shown.}
\label{fig:Jd1}
\end{figure}
As mentioned in the introduction,
when $J_{\times}\ll J_{\perp}$, the system adopts the
N-type state with ferromagnetic spin arrangements along the 
diagonals. When $J_{\times}\gg J_{\perp}$ the system adopts 
the R-type state. In this case, 
the AF spin arrangements shift to the diagonals, 
and spins on the rungs are forced to adopt a ferromagnetic 
arrangement. When $J_{\times}\gg J$ and $J_{\perp}\gg J$, 
the system adopts the F-type state, 
where the spins on the chains adopt a ferromagnetic arrangement 
and an AF one on the rungs.
We use the spin arrangements on the right panel in 
Fig.~\ref{fig:Jd1}, where the lattice is subdivided into two sublattices,
as a starting point. 
The Hamiltonian density for the N-type state is given by
\begin{eqnarray}
\mathcal{H} =
\begin{pmatrix}
-h' & iJ_{1}\sin k  & J_{\times1}\cos k & \frac{J_{\perp 1}}{2} \\
-iJ_{1}\sin k & -h' & \frac{J_{\perp 1}}{2} & J_{\times1}\cos k \\
J_{\times1}\cos k &  \frac{J_{\perp 1}}{2} & -h' & iJ_{1}\sin k \\
\frac{J_{\perp 1}}{2} & J_{\times1}\cos k & -iJ_{1}\sin k & -h'
\end{pmatrix},
\end{eqnarray}
where 
\begin{eqnarray}
J_{1} &=&  J(1+2Q), \nonumber \\
J_{\perp1} &=&  J_{\perp}(1+2P), \nonumber \\
J_{\times1} &=&  J_{\times}(1+2P'),\nonumber \\
h'&=& h - (2J + 2J_{\times} + J_{\perp})M_{z}.
\end{eqnarray} 
Diagonalizing $\mathcal{H}$ yields the following four eigenenergies 
\begin{eqnarray}
E_{N1}(k) \!&\! =\! -h'+&\! J_{\times1}\cos k 
+ \sqrt{J_{1}^{2}\sin^{2} k + \frac{J_{\perp 1}^{2}}{4}}, \nonumber \\
E_{N2}(k) \!&\! =\! -h'-& \!J_{\times1}\cos k 
+ \sqrt{J_{1}^{2}\sin^{2} k + \frac{J_{\perp 1}^{2}}{4}},\nonumber \\
E_{N3}(k) \!&\! =\! -h'+& \!J_{\times1}\cos k 
- \sqrt{J_{1}^{2}\sin^{2} k + \frac{J_{\perp 1}^{2}}{4}},\nonumber \\
E_{N4}(k) \!&\! =\! -h'-& \!J_{\times1}\cos k 
- \sqrt{J_{1}^{2}\sin^{2} k + \frac{J_{\perp 1}^{2}}{4}}. 
\end{eqnarray}
Similarly, the eigenenergies for the F-type state and for 
the R-type state are given respectively by
\begin{eqnarray}
E_{F}(k) \!&\! =\! &\! -h' \pm J_{1}\cos k \pm \sqrt{J_{\times1}^{2}\sin^{2}k 
+ \frac{J_{\perp 1}^{2}}{4}}, \nonumber \\
E_{R}(k) \!&\! =\! &\! -h' \pm \frac{J_{\perp 1}}{2}
\pm \sqrt{J_{1}^{2}\sin^{2}k 
+ J_{\times1}^{2}\cos^{2}k}.
\end{eqnarray}
The free energy per site is
\begin{equation}
F  =  JQ^2 + \frac{J_{\perp}P^{2}}{2} + J_{\times}P'^{2} 
+\frac{h}{2}-(J + J_{\times} + \frac{J_{\perp}}{2})M_{z}(M_{z}+1) 
- \frac{k_{B}T}{4N} \sum_{k}\sum_{p=1}^{4}\ln[1 + e^{-\beta E_{\kappa p}(k)}],
\end{equation}
where $\kappa\equiv N$, $R$, or $F$ depending on the state considered. 
$E_{\kappa p}(k)$ designates one of the four eigenenergies in state $\kappa$. 
The magnetization can be calculated using $M_{z} = -\pderiv{F}{h}$, which gives
\begin{eqnarray}
\label{eq:mag}
M_{z} =  -\frac{1}{2} +\frac{1}{4N}\sum_{k}\sum_{p=1}^{4} 
n_{F}[E_{\kappa p}(k)].
\end{eqnarray}
This adds another equation to the set of self-consistent equation 
satisfied by the spin bond parameters $Q$, $P$, and $P'$, 
which are obtained by minimizing the free energy:
\begin{eqnarray}
\label{eq:sc}
Q & = & -\frac{1}{8NJ}\sum_{k}\sum_{p=1}^{4}
\pderiv{E_{\kappa p}(k)}{Q}n_{F}[E_{\kappa p}(k)], \nonumber  \\
P & = & -\frac{1}{4NJ_{\perp}}\sum_{k}\sum_{p=1}^{4} 
\pderiv{E_{\kappa p}(k)}{P}n_{F}[E_{\kappa p}(k)],  \nonumber \\
P' & = & -\frac{1}{8NJ_{\times}}\sum_{k}\sum_{p=1}^{4} 
\pderiv{E_{\kappa p}(k)}{P'}n_{F}[E_{\kappa p}(k)].
\end{eqnarray}
The equations in (\ref{eq:sc}) are solved numerically in order 
to obtain the free energies, magnetization, bond parameters, 
and uniform spin susceptibility
as functions of magnetic field and temperature.

\section{Results}
\label{sec:results} 

\subsection{Free energies}
\begin{figure}
\subfigure{
\includegraphics[scale=0.2]{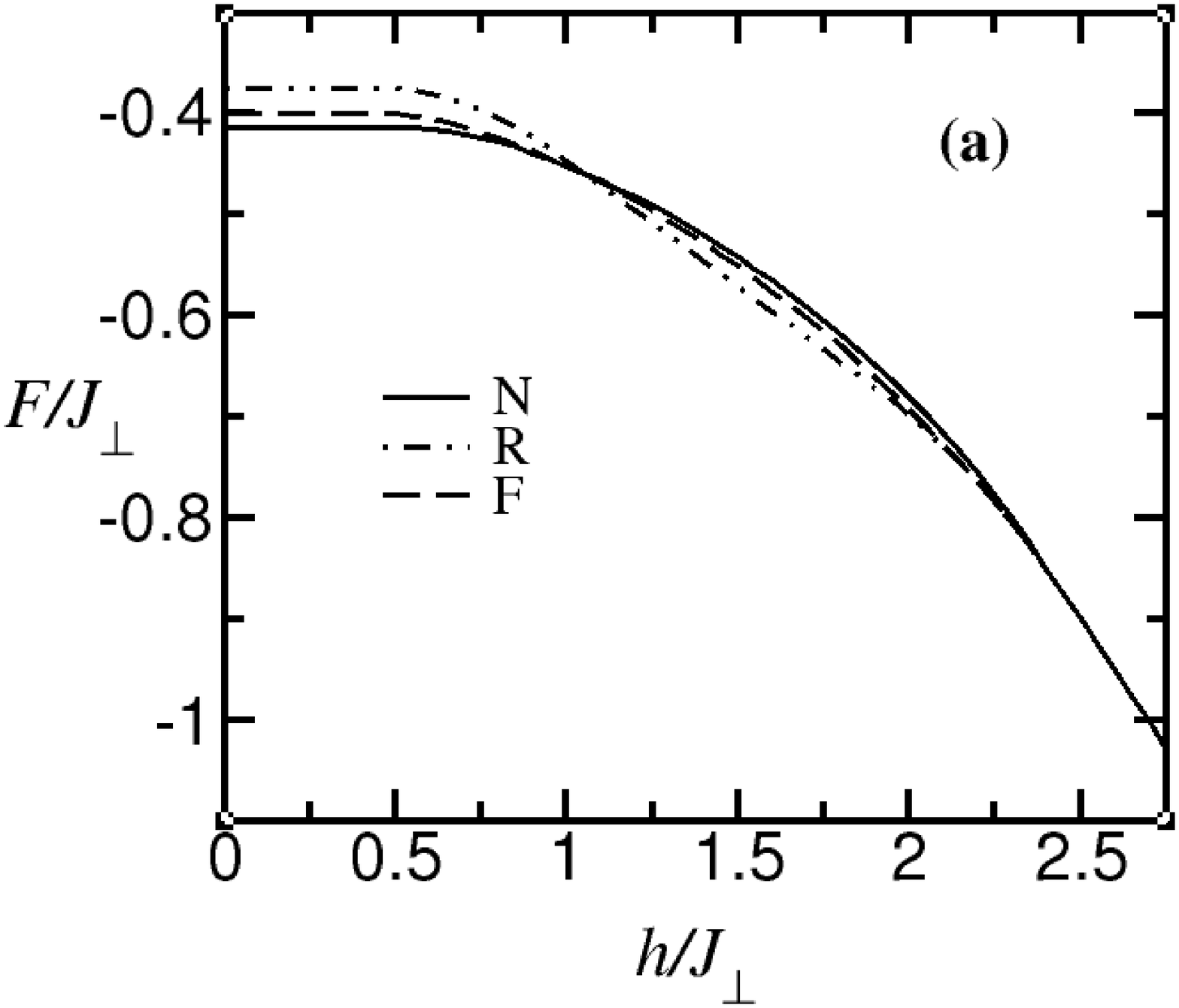}}
\subfigure{
\includegraphics[scale=0.2]{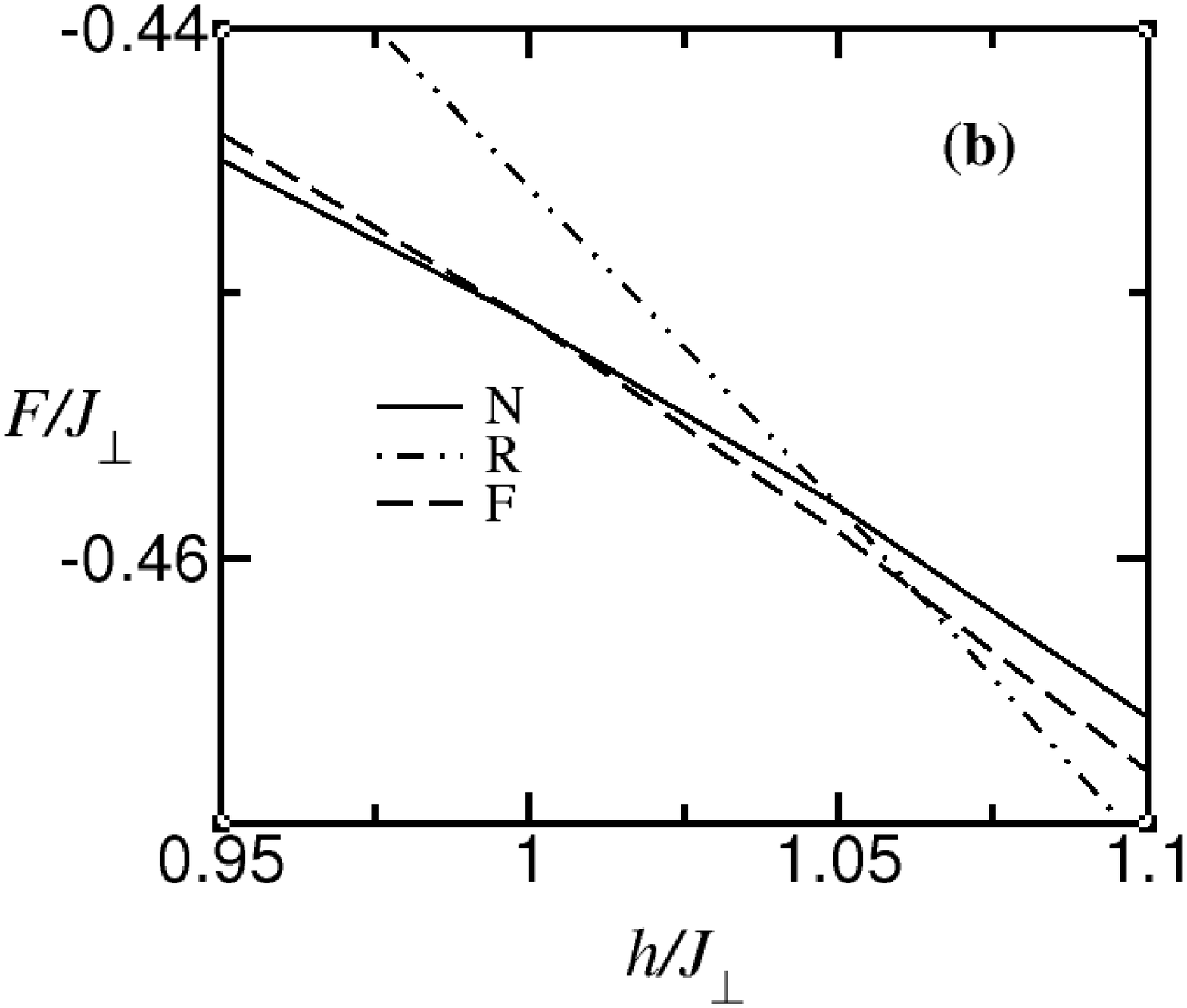}}
\centering
\caption{(a) The free energies are plotted as functions of magnetic 
 field for $J=0.5J_{\perp}$, and $J_{\times}=0.4J_{\perp}$. 
(b) The field-induced transitions are shown by zooming 
in on fields near the critical fields.}
\label{fig:Magfree}
\end{figure}

The free energies are compared as seen in Fig.~\ref{fig:Magfree} 
to determine the state with the lowest free energy, which is in turn used 
to obtain the parameters and magnetization.
As Fig~\ref{fig:Magfree}b shows for the given coupling constants set, 
a first transition occurs
from the N-type state to the F-type state, then another 
one from the F-type state to the R-type state.
Note that this couple of transitions 
was not mentioned in Ref.~\cite{Jd4}, which 
used the Lanczos method.
Our approach works 
well for analyzing the effect of magnetic field when 
$J_{\times}=0$~\cite{Azz5}, and when $h=0$ and 
$J_{\times}\neq 0$~\cite{ramakko2007}. 
Including both $J_{\times}$ and $h$ may be pushing the 
the current mean-field approach to its limits. In the strong field regime
all free energies are equal because the magnetization 
has saturated and all AF fluctuations have disappeared; i.e.,
even short-range AF correlations are absent in the 
saturated ferromagnetic state.
\begin{figure}
\begin{center}
\subfigure{
\includegraphics[scale=0.2]{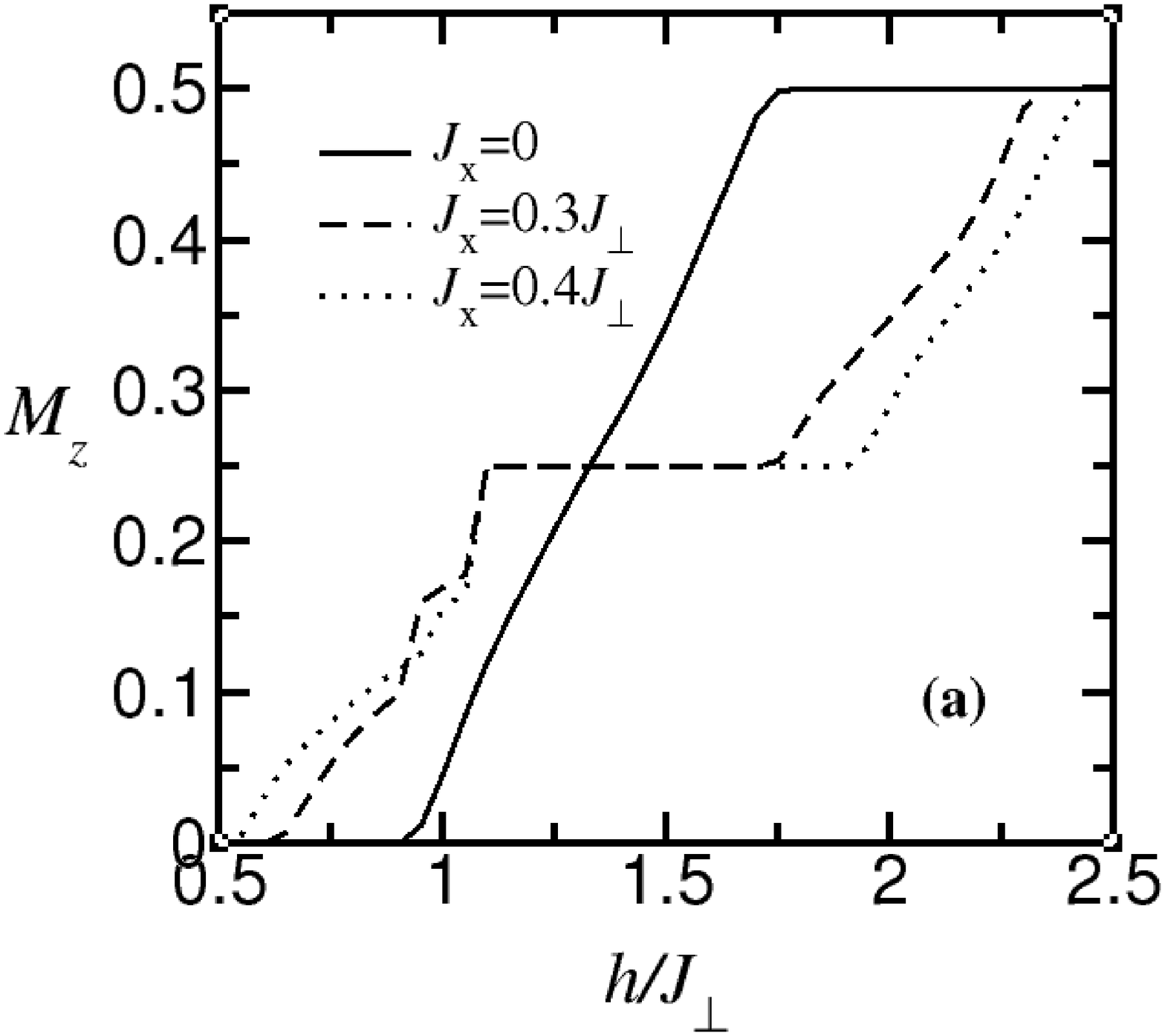}}
\subfigure{ 
\includegraphics[scale=0.2]{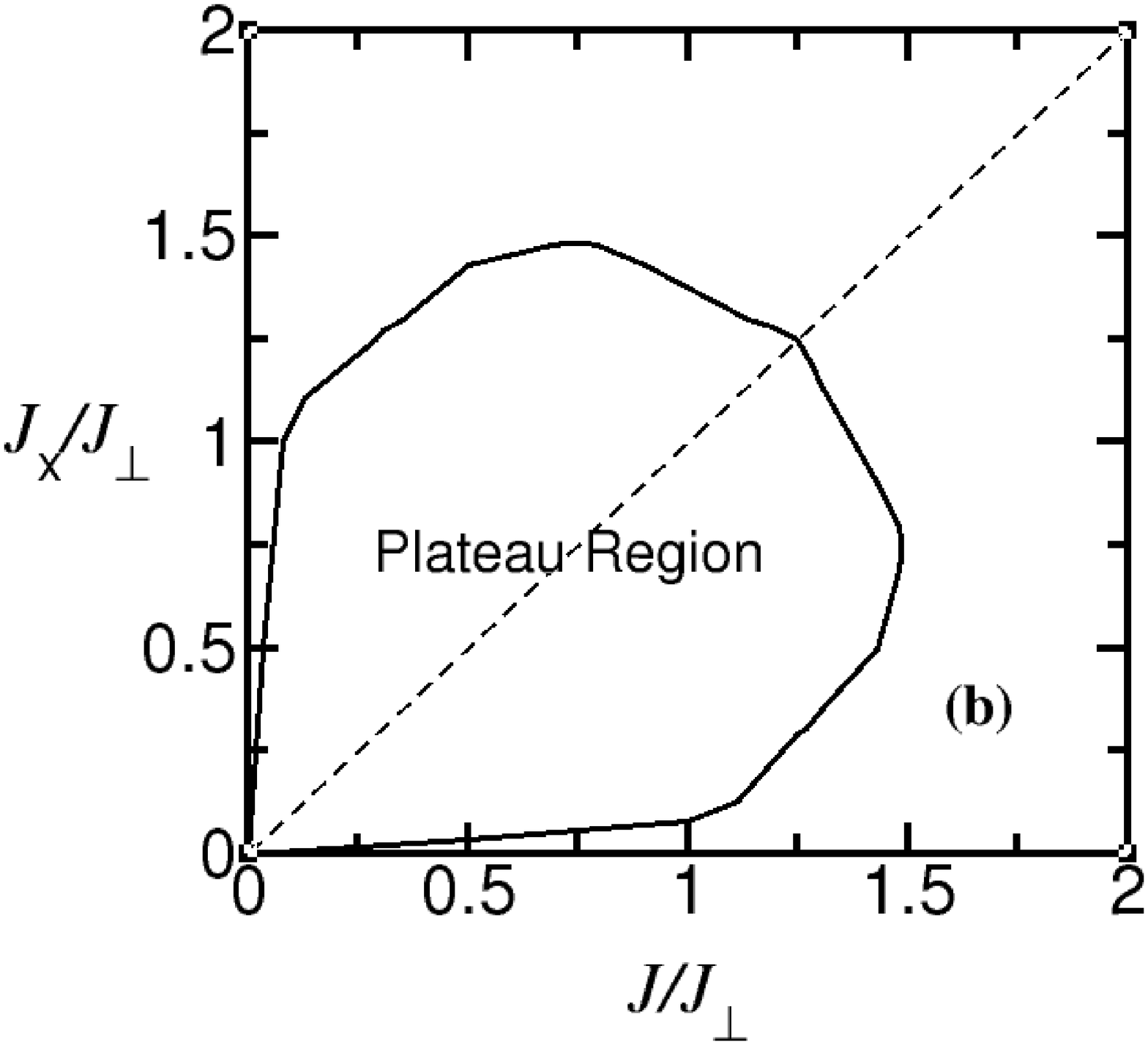}}
\subfigure{
\includegraphics[scale=0.2]{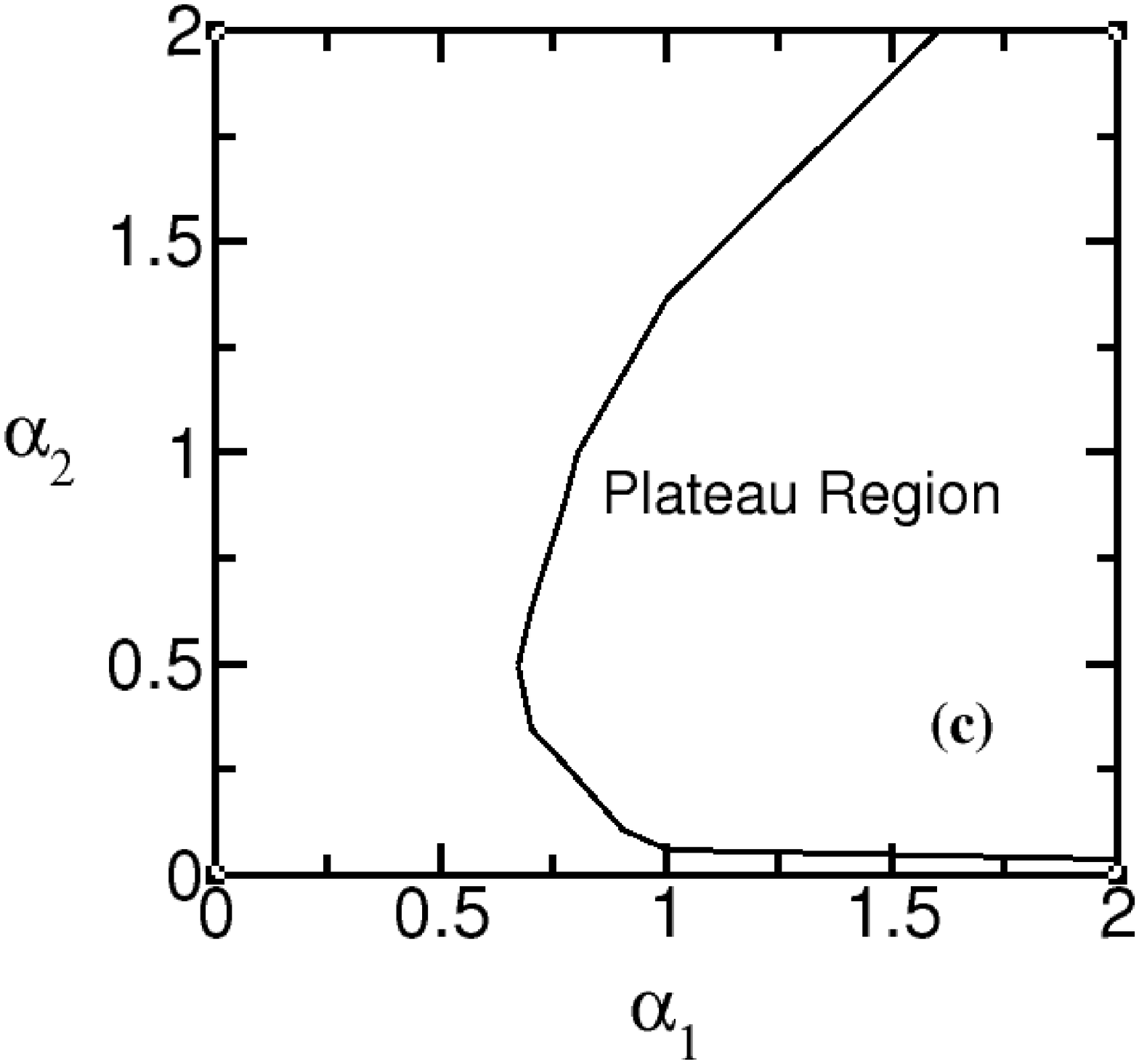}}
\caption{\label{fig:mag2} (a) The magnetization is plotted as
a function of $h/J_\perp$ for $J=0.5J_{\perp}$ 
and for the values of $J_\times/J_\perp$
given in the legend.
The $(J/J_{\perp},J_{\times}/J_{\perp})$ and $(\alpha_1=J_\perp/J,\alpha_2=J_\times/J)$ 
phase diagrams at $M_z=0.25$ are displayed in (b) and (c), respectively.}
\end{center}
\end{figure}

\subsection{Magnetization and phase diagrams}

The magnetization $M_z$ calculated using BMFT is displayed as a function 
of magnetic field in Fig.~\ref{fig:mag2}a for three sets of 
coupling values also used in Ref.~\cite{Jd4}. 
Because all the states are gapped in zero field, the magnetization 
remains zero for all fields smaller than the zero-field gap.
This is clearly observed in Fig. \ref{fig:mag2}a.
Our curves for $M_z$ compare very well qualitatively 
with existing numerical results~\cite{Jd4,Honecker2000}. For certain 
coupling values a plateau appears at $M_z=0.25$ which means that 
on average half the spins are aligned along the magnetic field. Remember 
that $M_z=0.5$ is the saturation value. The frustration stabilizes 
this state, and as $J_{\times}$ increases the size of the plateau 
also increases. The $(J/J_\perp,J_\times/J_\perp)$ 
and $(\alpha_1,\alpha_2)$
phase diagrams in Figs.~\ref{fig:mag2}b and \ref{fig:mag2}c 
show the region 
where the plateau at $M_z =0.25$ appears. 
Here, $\alpha_1=J_\perp/J$ and $\alpha_2=J_\times/J$.
This plateau region is 
larger than, and encompasses, the numerically calculated plateau 
region~\cite{Jd4}. We find that the plateau appears as 
$J_\times$ increases only when $J_\perp \geq 0.67J$. 
Also, when $J_\times >>J_\perp$ the plateau disappears. 
The sharp boundaries in the $M_z=0.25$ phase diagrams only appear at $T=0$.
At finite temperature they are replaced by crossovers.
Note that the $(J/J_\perp,J_\times/J_\perp)$ phase diagram
is symmetric with respect to the diagonal because
the Hamiltonian is symmetric under exchanging 
the $J$ and $J_\times$ terms \cite{ramakko2007}. 

The effect of temperature on the magnetization, and on the 
magnetization plateau
is illustrated in Fig.~\ref{fig:mag1}a. As temperature 
increases the size of the plateau decreases, and eventually disappears 
altogether at high enough temperature. 
Because of thermal excitations, spins originally
locked in the gapped state at zero temperature become available for 
alignment along the magnetic field. This leads to the linear increase of 
magnetization in the low-field regime for $T=0.3J_\perp/k_B$; 
Fig. \ref{fig:mag1}a.  At $T=0$ there is a critical 
field at which the magnetization reaches the saturation 
value $M_s=1/2$, but for nonzero temperatures the critical 
behaviour is replaced by a crossover regime in a way similar to 
what was found for the Heisenberg chain and two-leg ladder in the 
absence of frustration \cite{Azz5}.

\subsection{Spin bond parameters}

\begin{figure}
\includegraphics[height=2.0cm]{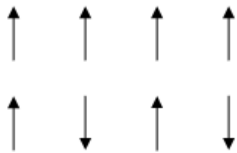}
\centering
\caption{An instantaneous configuration
where on average the 
magnetization per site $M_z=0.25$. For simplicity 
all the spins on one chain are ferromagnetically 
arranged while on the other chain they are antiferromagnetically arranged.}
\label{fig:mag4}
\end{figure}

\begin{figure}
\begin{center}
\subfigure{
\includegraphics[scale=0.2]{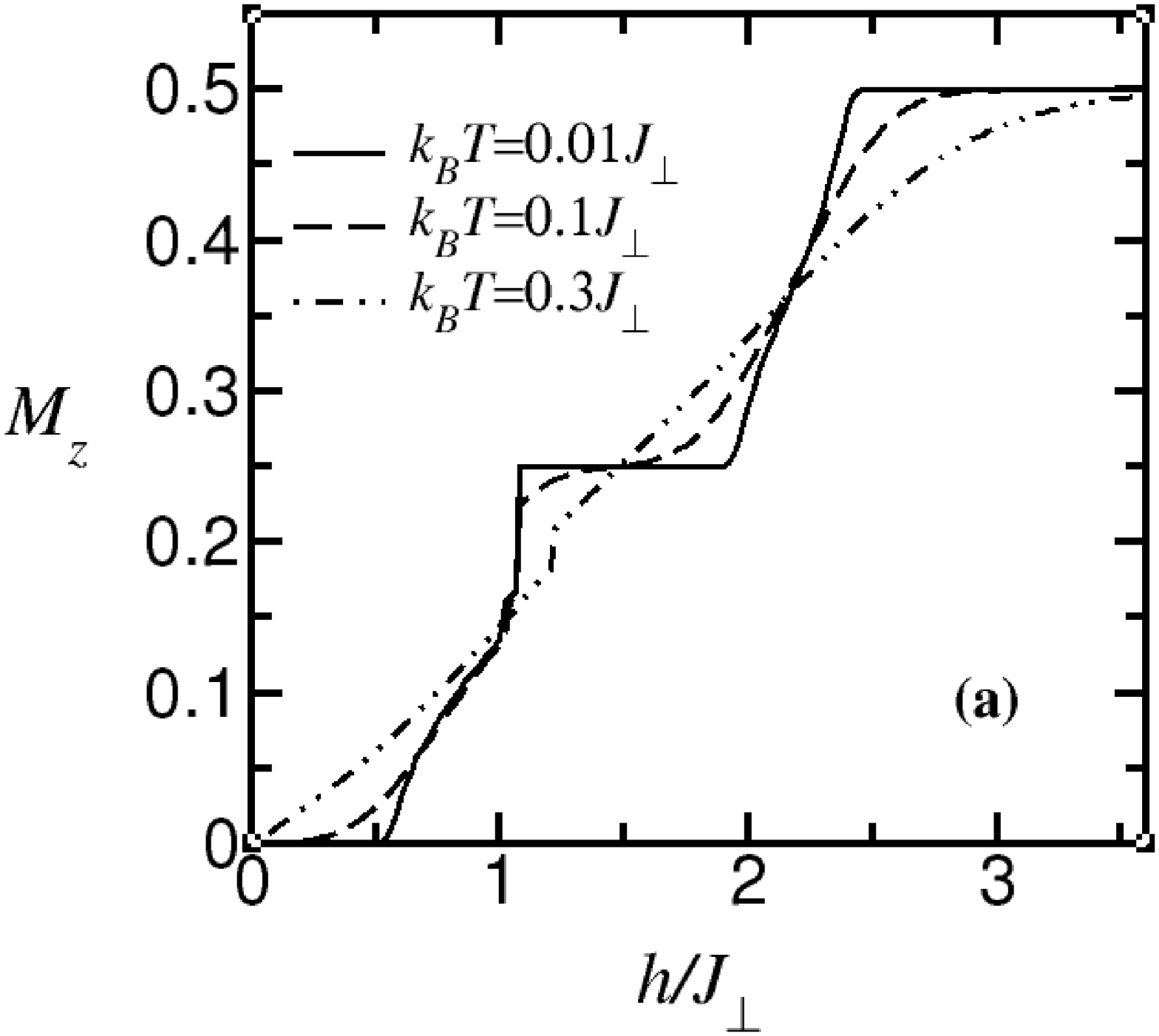}}
\subfigure{
\includegraphics[scale=0.2]{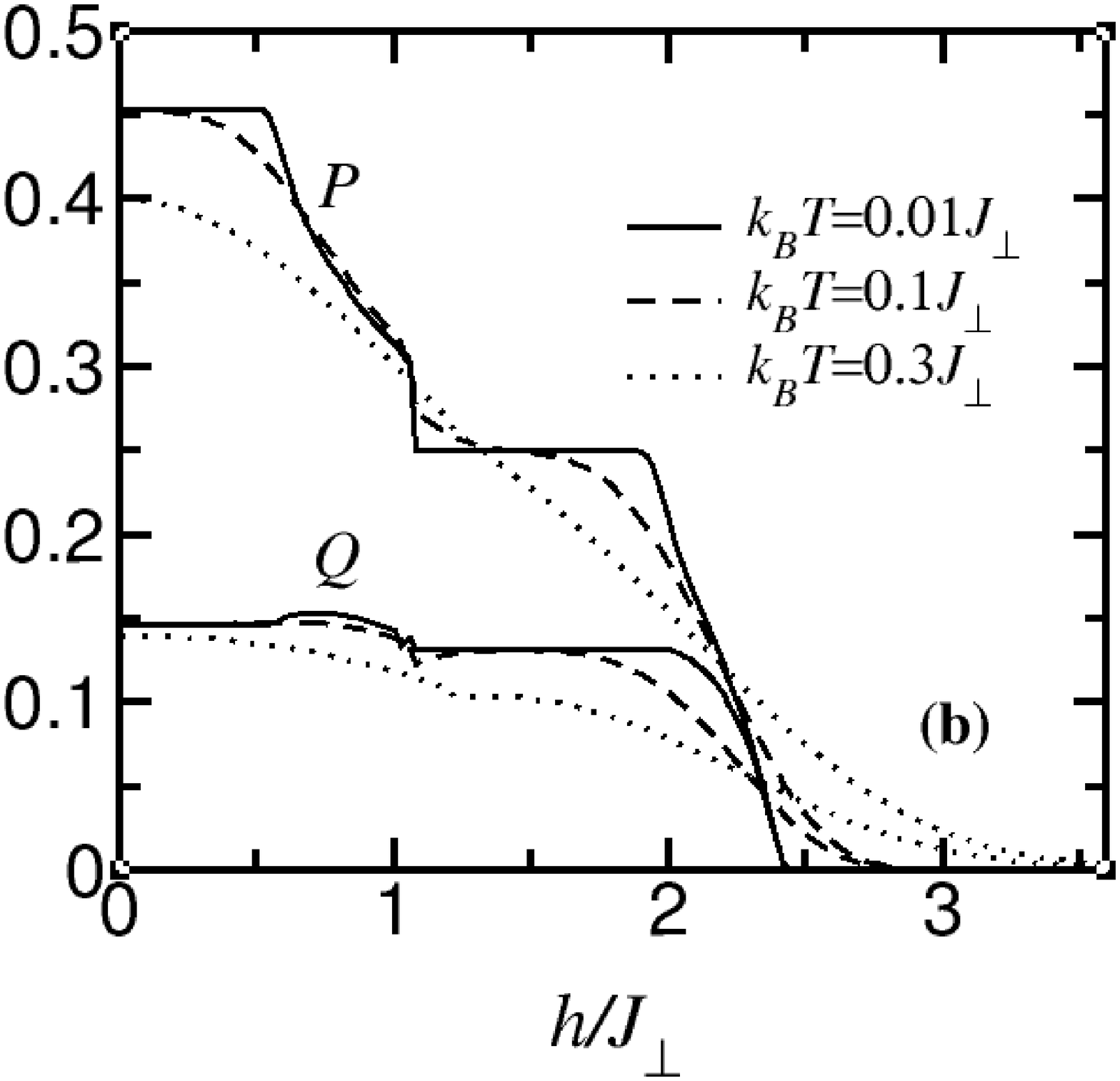}}
\subfigure{
\includegraphics[scale=0.2]{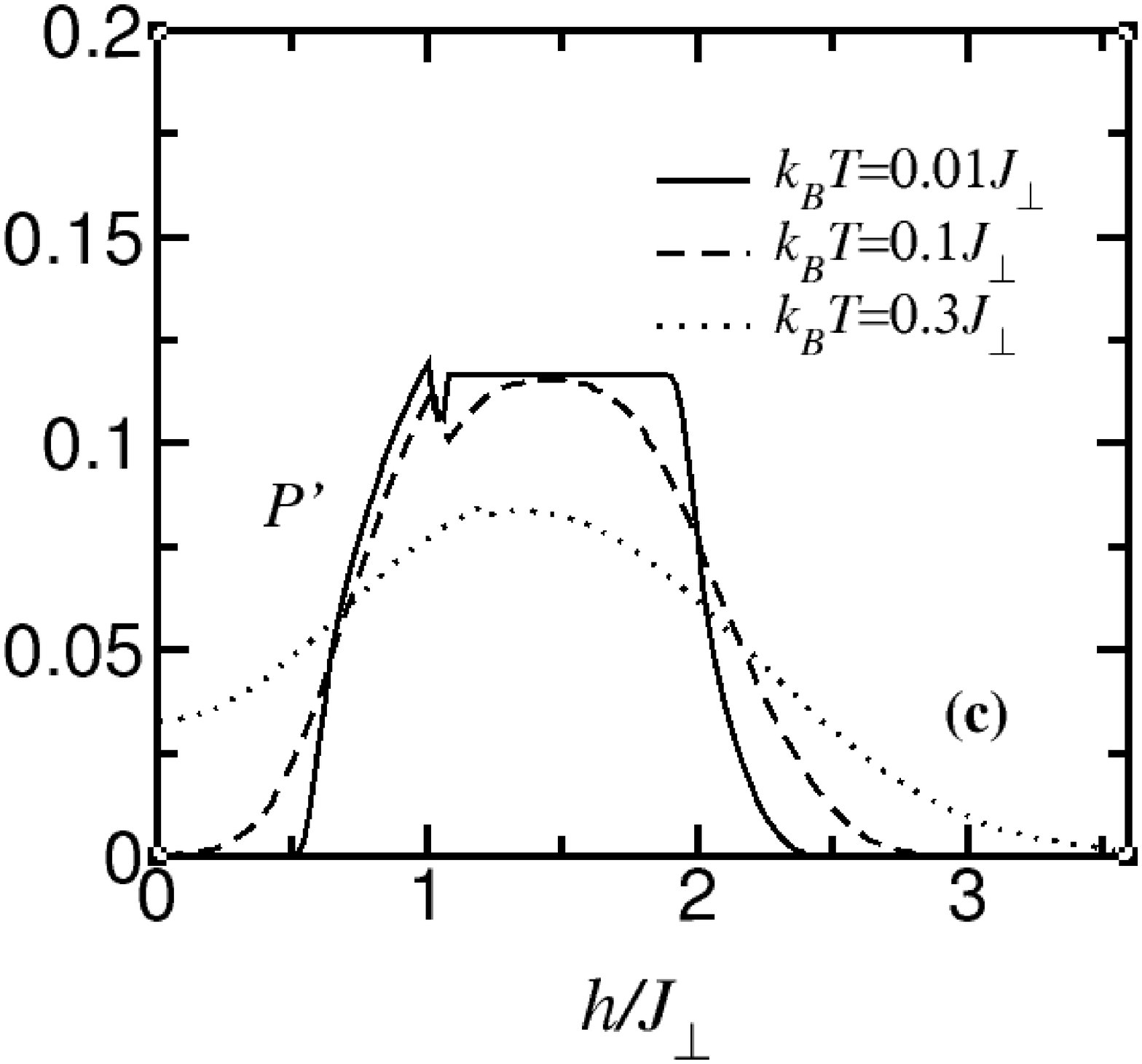}}
\caption{\label{fig:mag1} The magnetization in (a)  
and the spin bond parameters in (b) and 
(c) are plotted as functions 
of magnetic field for three different temperatures; 
$k_BT=0.01J_\perp$, $k_BT=0.1J_\perp$ and 
$k_BT=0.3J_\perp$. The exchange coupling constants are 
$J=0.5J_{\perp}$ and $J_{\times}=0.4J_{\perp}$.}
\end{center}
\end{figure}

The bond-mean-field parameters are shown in 
Figs.~\ref{fig:mag1}b and \ref{fig:mag1}c 
for $J=0.5J_\perp$ and $J_\times = 0.4 J_\perp$. 
At the plateau, statistically half of the spins have been aligned with 
the magnetic field because $M_z=0.25$. Surprisingly, the bond-mean-field 
parameters also adopt half of their maximum value. 
The maximum value of $Q$ and $P'$ is their value when $J_{\perp}=h=0$. 
The maximum value of $P$ is 0.5. 
For simplicity imagine an instantaneous configuration
where all the spins on one chain align along 
one direction, whereas the spin on the other chain which are 
antiferromagnetically arranged. This is illustrated in 
Fig.~\ref{fig:mag4}. In this case, the average magnetization 
per site is $M_z=0.25$, and
exactly half of the rung bonds, one of the chains, 
and half of the diagonal bonds are antiferromagnetically arranged. 
The bond parameters represent the average AF correlations in
the directions along which they are calculated, so each of them 
adopts half of its maximum value in zero field. The bond parameters 
vanish in the strong field regime, where the state labeling becomes 
obsolete and all three free energies equal.
The bond parameters vanish at the same field where 
the magnetization reaches saturation signaling in this way 
a quantum (zero-$T$) phase transition \cite{Azz5}.

When $T>0$, thermal fluctuations spoil this transition.
A crossover behaviour replaces the zero-$T$ critical behaviour
exactly in the same way as in the Heisenberg chain and non-frustrated
two-leg ladder \cite{Azz5}. At high enough temperature 
the plateaus in the bond parameters disappear; 
Figs. \ref{fig:mag1}b and \ref{fig:mag1}c.


\subsection{Uniform spin susceptibility}

\begin{figure}
\begin{center}
\subfigure{
\includegraphics[scale=0.2]{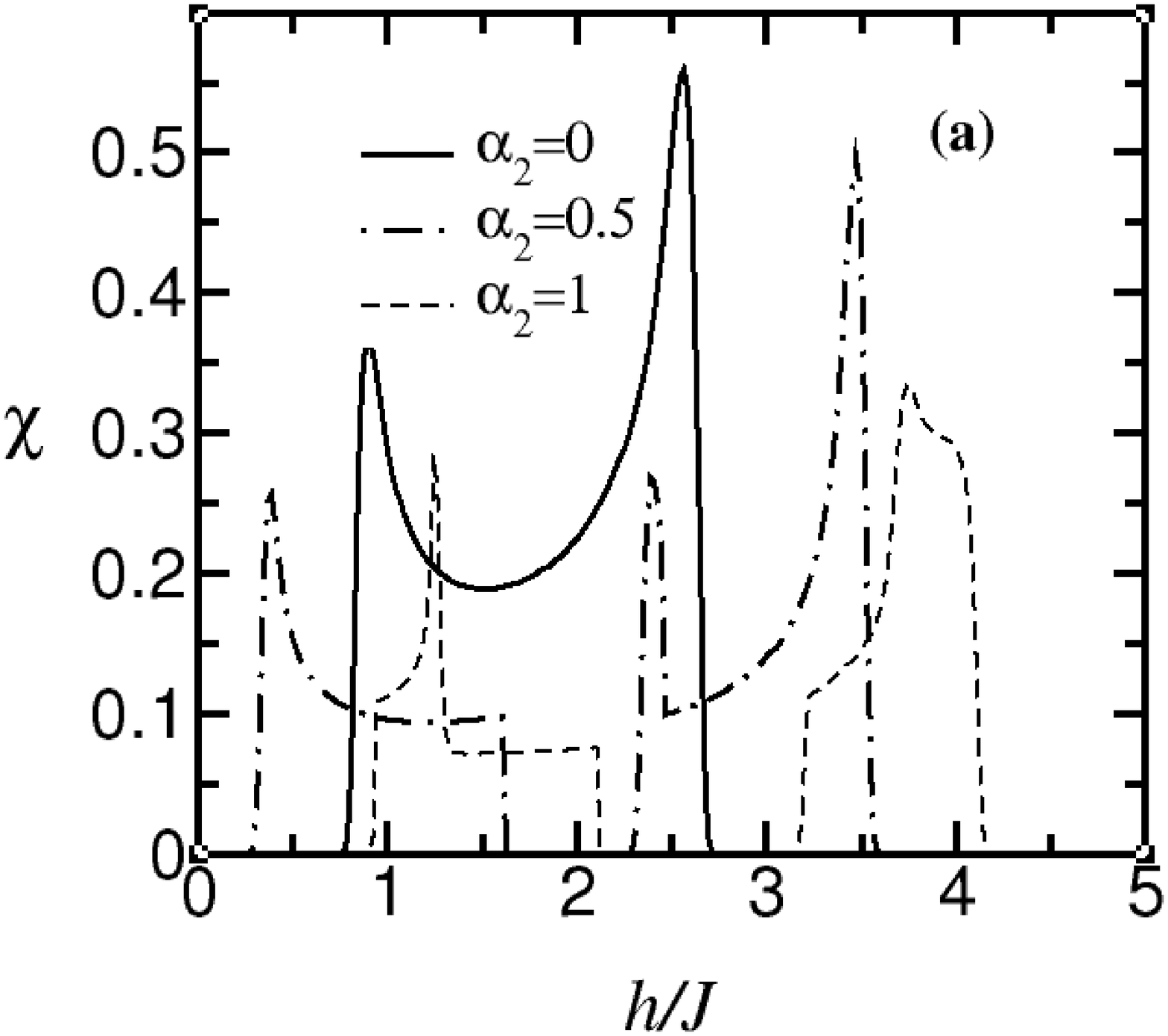}}
\subfigure{
\includegraphics[scale=0.2]{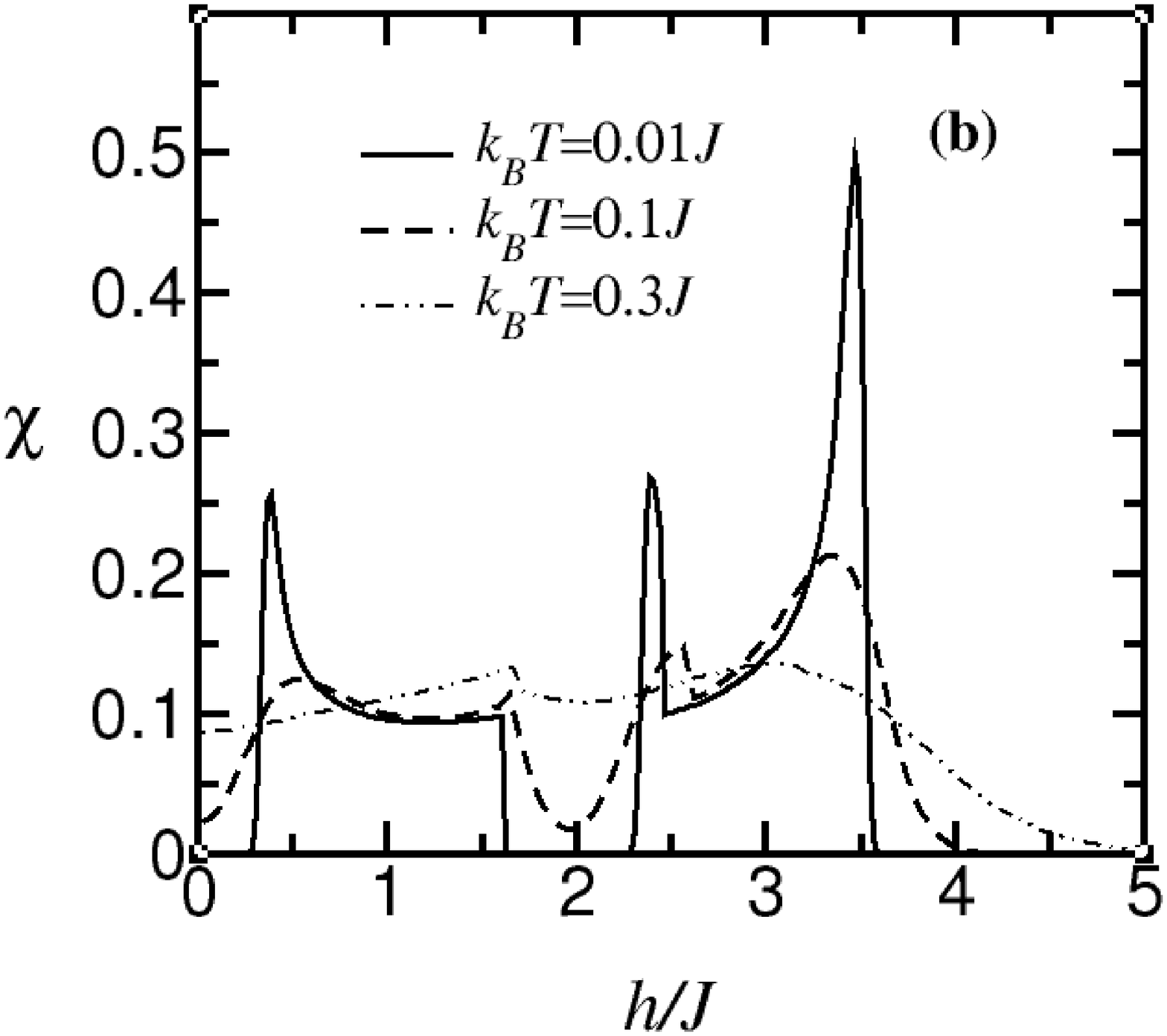}}
\caption{\label{fig:suscept} The susceptibility $\chi = \pderiv{M_z}{h}$ 
is plotted as a function of $h/J$ for 
(a) three coupling values at $k_BT/J=0.01$, and 
(b) three temperatures with $\alpha_1=1$ and $\alpha_2=0.5$.}
\end{center}
\end{figure}

For convenience, in this section the unit of 
energy is $J$ rather than $J_\perp$.
The magnetic susceptibility $\chi = \pderiv{M_z}{h}$
is shown as a function of field for three different couplings in 
Fig.~\ref{fig:suscept}a at temperature $k_BT/J=0.01J$, and for three 
different temperatures in 
Fig.~\ref{fig:suscept}b for $\alpha_1=J_\perp/J=1$
and $\alpha_2=J_\times/J=0.5$. 
For $\alpha_1=1$ and $\alpha_2=0$, $\chi$ is 
characterized by two peaks only, and is zero for low fields because 
of the energy gap. As $\alpha_2$ increases $\chi$ 
shows more structure and a gap.
The gap in $\chi$ is due to the plateau at $M_z=0.25$.
Thermal fluctuations round-off the peaks 
and close the gaps in  $\chi$ in agreement with 
the disappearance of the plateau at high enough temperatures.

\subsection{Interpretation of the results}

\begin{figure}
\begin{center}
\subfigure{
\includegraphics[scale=0.2]{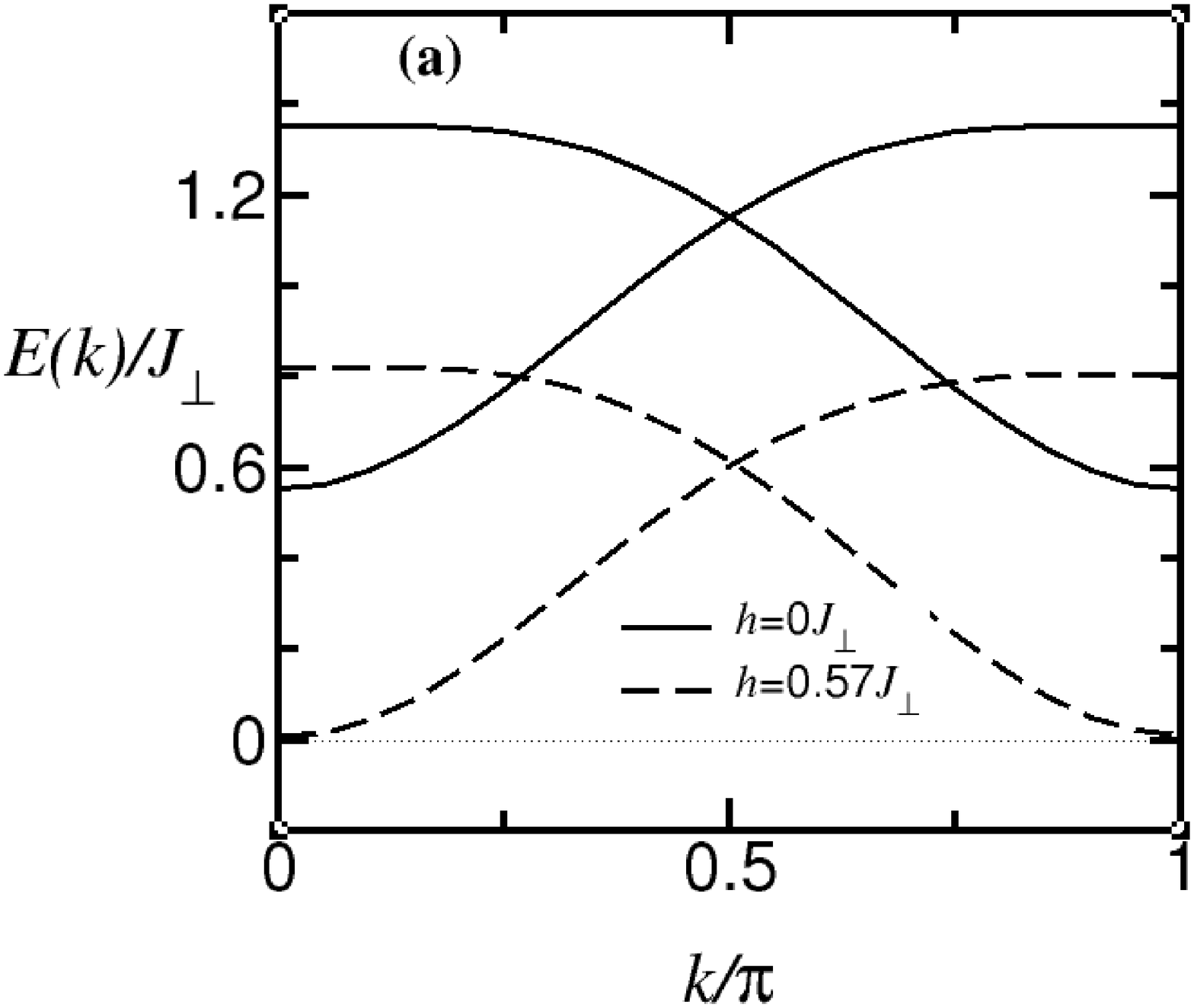}}
\subfigure{
\includegraphics[scale=0.2]{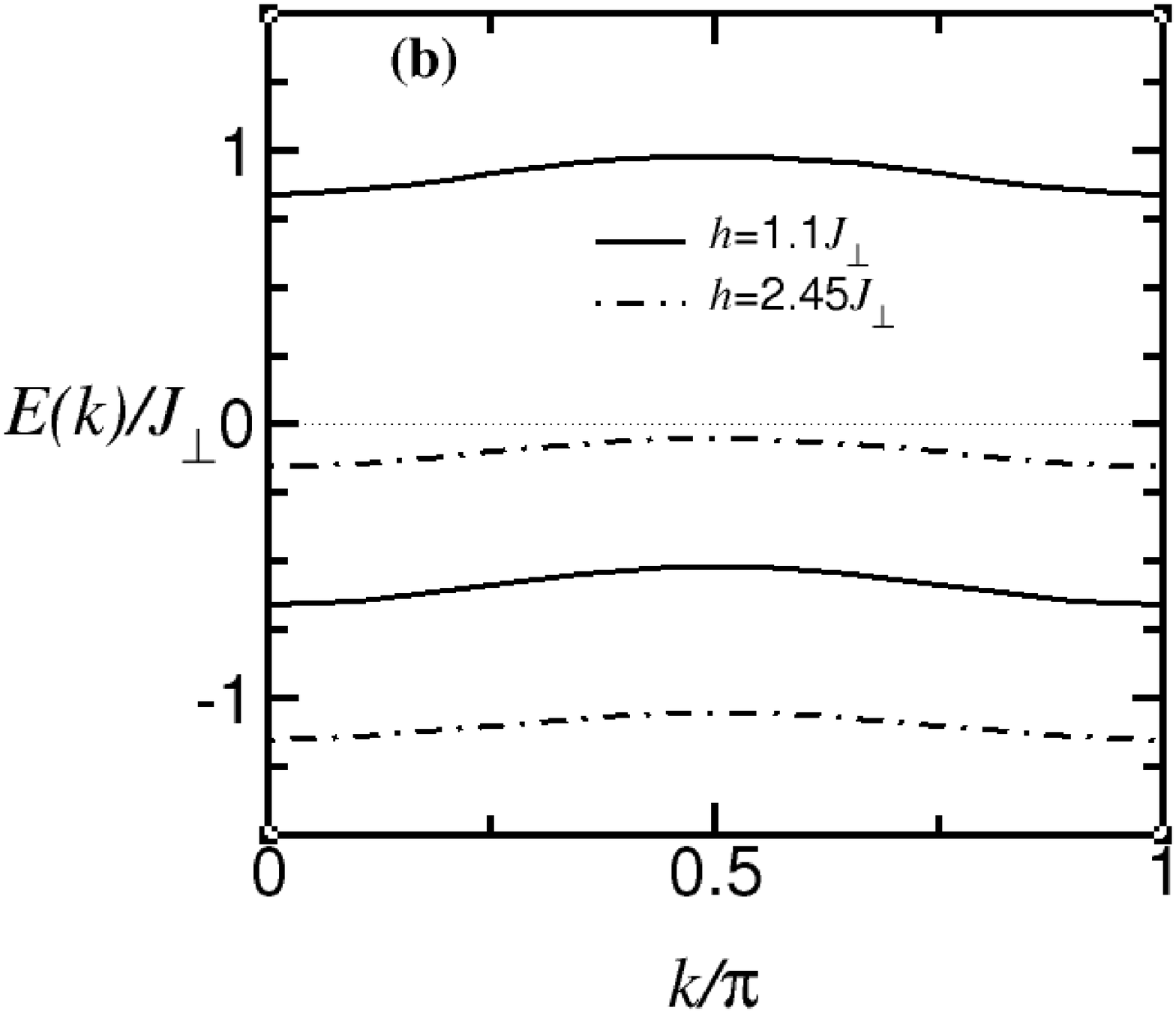}}
\caption{(a) The energies 
$E_{N1}$ and $E_{N2}$ are plotted as functions of wavenumber 
$k$ for $h=0J_{\perp}$ and $h=0.5J_{\perp}$. (b) The
energies $E_{R1}$ and $E_{R2}$ are plotted as functions of 
$k$ for $h=1.1J_{\perp}$ and $h=2.45J_{\perp}$. 
Here, $J=0.5J_{\perp}$, and $J_{\times}=0.4J_{\perp}$.
\label{fig:mag5} }
\end{center}
\end{figure}

The magnetization plateau and its behaviour in general 
can be explained using the chemical 
potential and energy band filling of the 
JW fermions in the same way as in 
Ref.~\cite{Azz5}. The excitation energies $E_{N1}$ and $E_{N2}$ 
are plotted in figure~\ref{fig:mag5}a as functions of wavenumber 
$k$ for two different values of field $h$, and for $J=0.5J_{\perp}$ and 
$J_{\times}=0.4J_{\perp}$. When $h=0$, the chemical potential of 
the JW fermions is zero; so only the lower-energy bands $E_{N3}$ 
and $E_{N4}$ are filled. A glance at Eq.~(\ref{eq:mag}) reveals 
that since two out of four bands are filled it is clear that $M_z=0$. 
As the field increases, the chemical potential of the JW fermions 
increases. For fields $h$ smaller than the energy gap, $M_z$ remains 
zero. For the present set of 
exchange coupling constants,  $h$ must be greater than
$0.57J_{\perp}$, which is the value of the energy gap in 
the absence of a magnetic field, in order 
to get a nonzero magnetization; i.e., 
for fields greater than this threshold 
value the population of the higher-energy bands increases as field 
increases, leading to an increasing magnetization $M_z$. At 
$h=1.1J_{\perp}$ the state of the system becomes the R-type state. 
The excitation energies $E_{R1}$ and $E_{R2}$ are plotted in 
Fig.~\ref{fig:mag5}b as functions of $k$ for two 
different values of field. The energy bands in the R-type 
state are separated by a gap for certain coupling values whereas in the 
other two states they are not. From $h=1.1J_{\perp}$ to 
$h=1.9J_{\perp}$ three of the bands are populated and the fourth 
is empty so that there is a plateau at $M_z=0.25$. Further increasing 
the field increases the population of the fourth band until it becomes 
completely full at $h=2.42J_{\perp}$. At this point $M_z=0.5$, and further 
increasing the field can no longer increase the population of 
the JW fermions. This results in the saturation of the magnetization $M_z$.

\section{Conclusion}
\label{sec:Conclusion}

In this work we analyzed the interplay between 
the quantum criticalities induced by frustration and magnetic field 
in the two-leg antiferromagnetic Heisenberg ladder.
When frustration is introduced in the presence of a magnetic field
a plateau appears at half the saturation
magnetization. The plateau in the magnetization appears because 
of the onset of an energy gap between two (excitation) energy 
bands. As frustration increases this plateau increases in size indicating 
that this phase is stabilized by the frustration. Our results 
agree qualitatively well with existing numerical data of Sakai and Okazaki.
In the presence of frustration, the system is characterized 
by three spin bond parameters, one along the chains, one along the rungs
and one along the diagonals. These parameters vanish in the strong field limit
signaling the occurrence of a zero-$T$ phase 
transition. The latter give place to a crossover behaviour when temperature
becomes nonzero.
The magnetization plateau shrinks as temperature increases, and disappears
in the high-$T$ regime. The spin-bond parameters are also 
characterized by a plateau that disappears in the high-$T$ regime.

\section*{Acknowledgments}
We wish to acknowledge the financial support of the Natural Science 
and Engineering Research Council of Canada
(NSERC), and of the Laurentian University Research Fund (LURF).


\begin{thebibliography}{5}
\bibitem{ramakko2007}B. Ramakko and M. Azzouz, Phys. Rev. B {\bf 76}, 064419 (2007).

\bibitem{Jd4}T. Sakai and N. Okazaki, Journal of Applied Physics \textbf{87}, 5893 (2000).

\bibitem{Honecker2000}A. Honecker, F. Mila, and M. Troyer, 
Eur. Phys. J. B \textbf{15}, 227 (2000).











\bibitem{Azz1}M. Azzouz, Phys. Rev. B \textbf{48}, 6136 (1993).

\bibitem{Azz2}B. Bock and M. Azzouz, Phys. Rev. B \textbf{64}, 054410 (2001).

\bibitem{Azz3}M. Azzouz, L. Chen, and S. Moukouri, Phys. Rev. B \textbf{50}, 6233 (1994).



\bibitem{Azz5}M. Azzouz, Phys. Rev. B \textbf{74}, 174422 (2006).


\bibitem{mermin1966}N.D. Mermin and H. Wagner, Phys. Rev. Lett. \textbf{17}, 1133 (1966).
\bibitem{JW}P. Jordan and E.Wigner, Z. Phys. \textbf{47}, 631 (1928).



\bibitem{affleck1988}I. Affleck and J.B. Marston, Phys. Rev. B \textbf{37}, 3774 (1988).
\bibitem{Cloiseaux}J. des Cloiseaux and J.J. Pearson, Phys. Rev. \textbf{128}, 2131 (1962).
























































\end{thebibliography}
\end{document}